# Skyrmion-based Leaky Integrate and Fire Neurons
# for Neuromorphic Applications

Aijaz H. Lone[1], Selma Amara[1], Fernando Aguirre [2], Mario Lanza[3] and H. Fariborzi[1]

[1]*Computer, Electrical and Mathematical Science and Engineering Division, King Abdullah University of Science and Technology, Saudi Arabia*
[2]*Dept. D'Enginyeria Electronica, Universitat Autonoma de Barcelona*

[3]*Physical Science and Engineering Division, King Abdullah University of Science and Technology, Saudi Arabia*

## ABSTRACT

Spintronics is an important emerging technology for data storage and computation. In this field, magnetic skyrmion-based devices are attractive due to their small size and energy consumption. However, controlling the creation, deletion and motion of skyrmions is challenging. Here we propose a novel energy-efficient skyrmion-based device structure, and demonstrate its use as leaky integrate (LIF) and fire neuron for neuromorphic computing. Here we show that skyrmions can be confined by patterning the geometry of the free layer in a magnetic tunnel junction (MTJ), and demonstrate that the size of the skyrmion can be adjusted by applying pulsed voltage stresses. A spiking neural network (SNN) made of such skyrmion-based LIF neurons shows the capability of classifying images from the Modified National Institute of Standards and Technology (MNIST) dataset.

*Index Terms*— Skyrmion, magnetic tunnel junction (MTJ), LIF neuron, spiking neural network (SNN), neuromorphic computing,

## INTRODUCTION

Magnetic skyrmions are topologically-protected swirling magnetic textures[12] with diameter down ~1 nm [3] that can be moved laterally by very small current densities $10^6$ A/m$^2$ [4], while their counterpart domain walls normally require $10^{12}$ A/m$^2$. Owing to these properties, magnetic skyrmions are being highly explored for high-density and low energy consumption data storage, and proof-of-concept devices for race track memory and computing [5], such as logic gates [6] and transistors [7], have been demonstrated. Additionally, magnetic skyrmion-based devices are showing potential for unconventional computing, including neuromorphic [8], reservoir [9] and reversible computing [10].

Skyrmions can be generated in some non-centrosymmetric magnetic compounds (such as MnSi [11]) and systems (such as Pt/Co/Ta [12] and Pt/CoFeB/MgO [13], among others). However, the fabrication of skyrmion-based devices require accurate control of their creation and deletion processes [14], which is challenging given their small size. Moreover,  governing the  skyrmions' motion in the device is challenging because the controlling current must be above the de-pinning level, but at the same time it should be low enough to prevent distorting the skyrmion shape, which would cause a non-linear behavior [15].

In order to mitigate some of these challenges, we present a skyrmion-MTJ device and its application in information storage and neuromorphic computing. We propose a top-down CoFeB/MgO/CoFeB/Ta structure in which the thickness of the bottom (free) CoFeB layer is adjusted strategically. In such structure, the skyrmions in the free layer are confined in the space within squared thicker regions as they create magnetic energy wells. First, we successfully test the stability of the confinement in presence of large spin transfer torque (STT) and spin orbit torque (SOT) currents. We control the size of the skyrmions by applying pulsed voltage stresses, which vary the surface anisotropy at the interface between the heavy metal layer and the free layer, thus modulating the skyrmion size. We observe that these devices can be operated as a Leaky-Integrate-and-Fire (LIF) neuron, and demonstrate that they can be employed for the construction of spiking neural networks (SNNs) capable of classifying the images from the MNIST dataset.

## VOLTAGE CONTROLLED SKYRMION DEVICES

We propose and simulate a neuromorphic system based on a matrix of skyrmion-MTJ devices (CoFeB/MgO/CoFeB/Ta), using MuMax and MATLAB. A schematic is shown in Fig. 1(a). The thickness of the CoFeB free layer (FL) is reduced from 1 nm to 0.8 nm for 20 nm every 100 nm to confine the skyrmions in areas of 100 nm × 100 nm. Such structure could be easily fabricated using standard electron beam lithography and etching processes. Hence, FL thickness ($t_{FL}$) in the 20-nm-wide spacer regions (0.8 nm) is smaller than in the 100-nm-wide regions (1 nm). The total anisotropy energy density ($K$) is given by [16]:

$$K = K_B + \frac{K_I}{t_{FL}} \qquad (1)$$

where $K_B$ is the bulk anisotropy energy density and $K_I$ is the interface anisotropy energy density. Thus, the anisotropy energy density of the 20-nm-wide regions is higher than that of the 100-nm-wide regions. This creates a magnetic energy well for confinement of the skyrmions. In order to minimize the tunneling current via MTJ to ensure that the STT doesn't affect the skyrmion size, we consider a 3nm MgO tunnel barrier (TB) in our device structure shown in Fig. 1(a). The magnetization of the free layer can be tuned by changing the diameter of the skyrmion by applying voltage pulses at the TB/FL interfaces as shown in Fig.1(b). The whole structure is simulated in MuMax and the magnetization of one skyrmion-based device and its correlation with the height of the applied voltage pulses is plotted in Fig. 1(c)). The role of DMI and anisotropy on magnetization switching is shown in Fig. 1(d) shows the magnetization switching for different anisotropy and DMI values. We observe that the VCMA switching behavior depends on the value of the DMI coefficient.

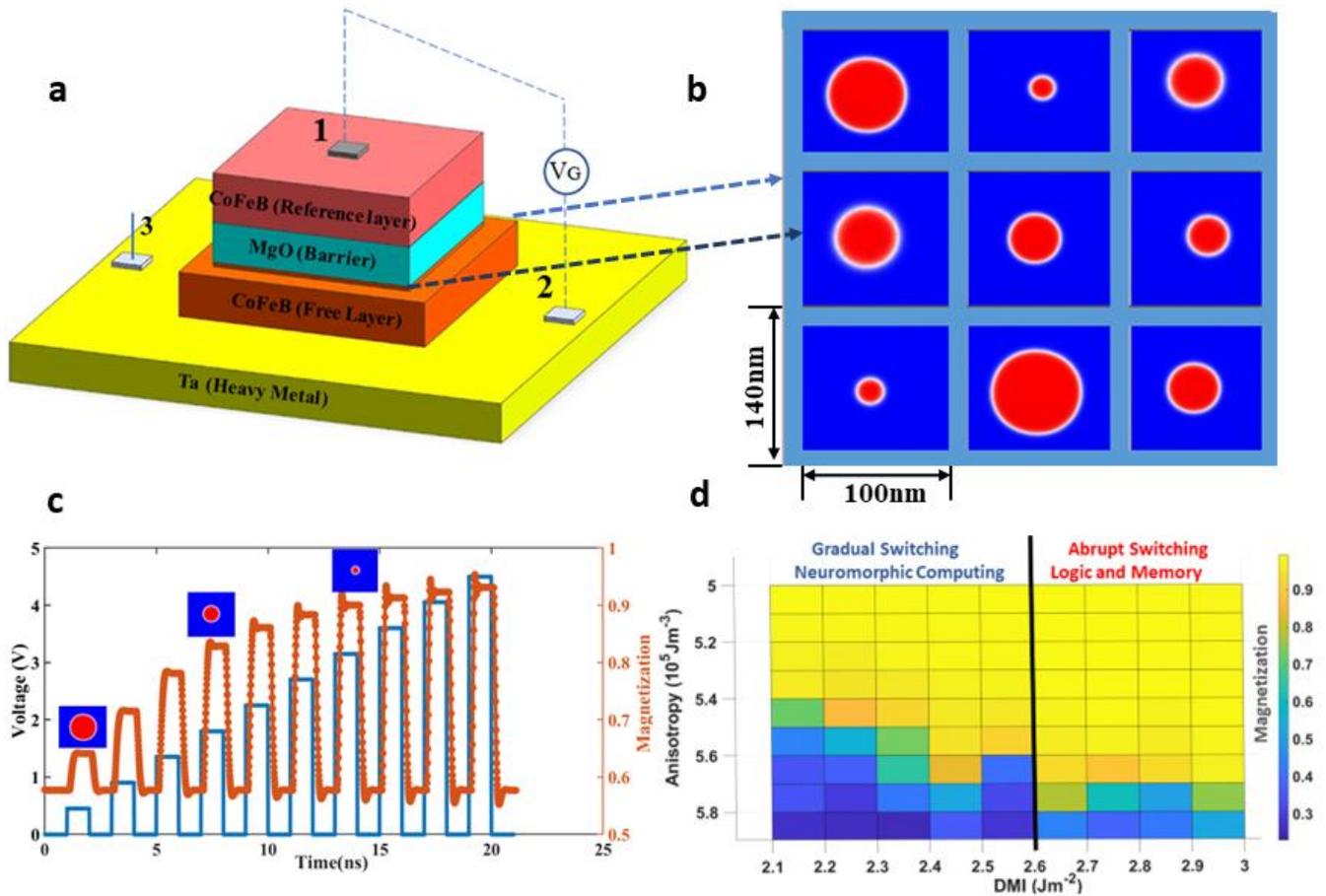

**Fig. 1. (a)** Schematic of the proposed matrix of 100nm × 100nm skyrmion-based devices, which is based on an MTJ. the dark region represents the free layer magnetization with skyrmions confined by brighter (high anisotropy) regions. **(b)** Voltage controlled skyrmion diameter. **(c)** voltage control of free layer magnetization by tuning the skyrmion size. **(d)** Normalized magnetization switching characteristics with respect to the DMI and the anisotropy. The results reveal different regions for device applications.

We can have either abrupt switching for logic and memory applications for values (DMI > 2.6 mJ.m$^{-2}$), In this region a 1.5% anisotropy change abruptly switches magnetization from 0.4 to 0.9. For DMI <2.6 mJ.m$^{-2}$ switching is discrete but gradual, and thus more suitable for neuromorphic computing

## SKYRMION MTJ DEVICE MODELING

The magnetic skyrmions are defined by their topological number, often named skyrmion number (Q) and calculated as follows [17]:

$$Q = \frac{1}{4\pi} \int \int \boldsymbol{m}.\left(\frac{\partial \boldsymbol{m}}{\partial x} \times \frac{\partial \boldsymbol{m}}{\partial y}\right) dx dy \qquad (2)$$

The spins are projected on the x-y plane and $\boldsymbol{m}$ is the normalized magnetization vector. The total magnetic energy of the free layer includes the exchange energy, Zeeman energy, uniaxial anisotropy energy, demagnetization energy and DMI energy [18]:

$$E(\boldsymbol{m}) = \int_V [\, A(\nabla \boldsymbol{m})^2 - \mu_0 \boldsymbol{m}. H_{ext} - \frac{\mu_0}{2} \boldsymbol{m}. H_d - K_u(\hat{u}. \boldsymbol{m}) + \boldsymbol{\varepsilon_{DM}}\,] dv \qquad (3)$$

where, $A$ is the exchange stiffness, $\mu_0$ is the permeability, $K_u$ is the anisotropy energy density, $H_d$ is the demagnetization field, and $H_{ext}$ is the external field. The DMI energy density can be computed by:

$$\varepsilon_{DM} = D[m_z(\nabla. \boldsymbol{m}) - (\boldsymbol{m}. \nabla). \boldsymbol{m}] \qquad (4)$$

We add SOT as a custom field term in MuMax[19].

## RESULTS AND DISCUSSION

Fig. 2(a-b) show the position and speed of the skyrmion (respectively) when a pulse voltage of 0.36V is applied between terminals 2 and 3 (this generates a current density in-plane along the heavy metal layer of $1 \times 10^{12}$ A/m², which generates the SOT). Normally without confinement the skyrmion would move smoothly along $X$ axis and also gets deflected in $Y$ direction, due to skyrmion Hall effect. However, On the contrary, for the confined case we observed that the skyrmion movement is restricted to 1 nm in the x direction and 6 nm in the y axis, even for even for a current density $J$ of $1 \times 10^{12}$ A/m² (see Fig. 2a)". This is further proved by the skyrmion velocity plot shown in Fig. 2(b). The velocity in absence of current (i.e., absence of SOT) is constantly zero; when a current pulse is applied laterally via heavy metal to generate SOT the velocity spikes to 12 m/s but it gets damped to zero within 1.6 ns. Thus, the method provides a reliable confinement of individual skyrmions. When tuning the diameter of the skyrmion via voltage pulses, it is important that the STT generated by tunnelling current is very low, so that the diameter of the skyrmion is minimally affected and any resulting switching error is avoided. Fig. 32(c) shows the variation

in magnetization due to STT for varied angles between free layer (skyrmion) magnetization and fixed layer magnetization. Clearly maximum variations are seen for angle $\frac{\pi}{2}$. The inset Fig. 2(d) the perpendicular component of the STT ($\tau_X$) increases with increasing angle and is maximum for an angle $\frac{\pi}{2}$. Due to the reduced polarized tunnelling current, the STT error decreases with the increasing MgO thickness, and we find that a thickness ~3 nm is good enough for a reliable voltage-controlled magnetization anisotropy (VCMA) see Fig. 2(d). We combine the magnetization behaviour (calculated with MuMax) with the Non-equilibrium Green's Function (NEGF) [20]to compute (using MATLAB) the tunnelling current and STT on the free layer due to this tunnel current.

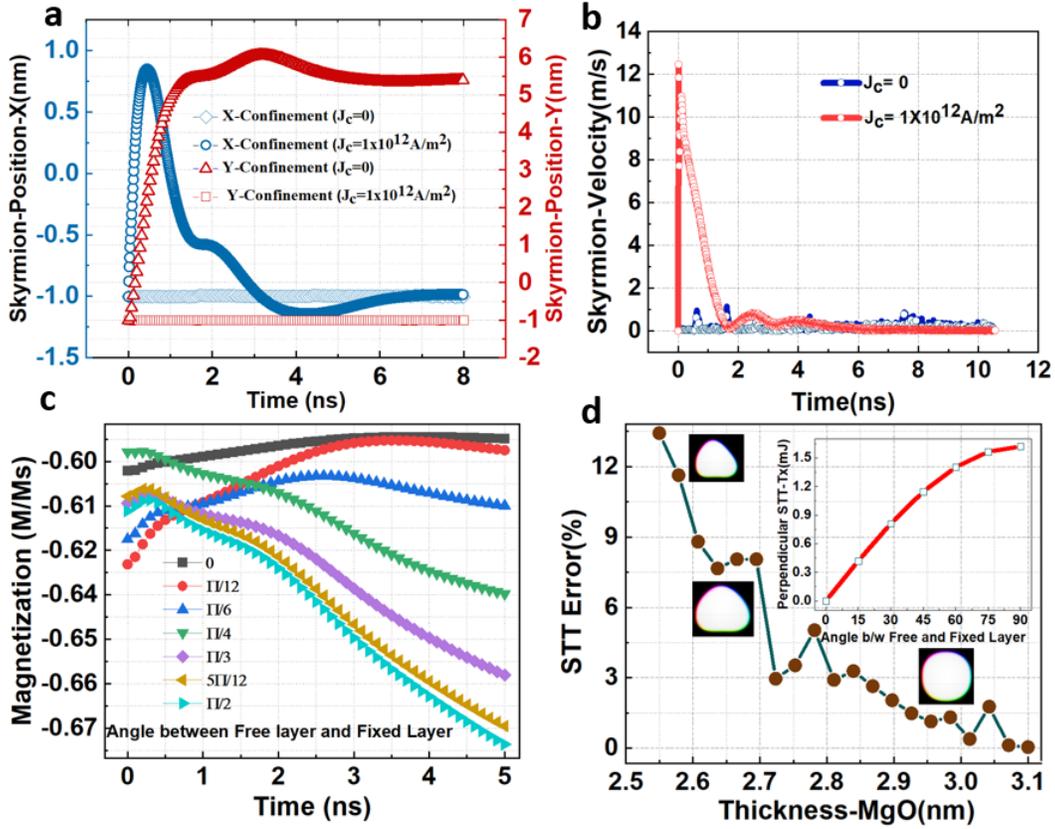

Fig. 2: (a) The reliable confinement of skyrmion in the magnetic energy well, skyrmion motion is restricted to just 6 nm in y and 2 nm in x direction. (b) Skyrmion velocity without/with the driving current, in case of a large current density skyrmion tries to gain momentum but the velocity drops abruptly due to confinement. (c) Effect of STT on free layer magnetization indicates max error for worst case (pi/2) is <12%. (d) Optimization of barrier thickness for mitigating worst case STT-error caused by tunnelling current and keeping a good RA product. The angle dependence of STT in units mJ is shown in the inset.

Using the external voltage, we tune the anisotropy and the DMI, which changes the skyrmion size; thus, the output of the MTJ is also modified. The dependence of the magnetic anisotropy (at the interface between the free layer and tunnel barrier layers) with the voltage can be expressed by [21].

$$K_S(V) = K_S(0) - \frac{\xi E}{t_{FL}} \qquad (5)$$

Where $K_S$ is the anisotropy, $E$ is the electric field at the MgO/CoFeB interface, $\xi$ is the VCMA coefficient and $t_{FL}$ is the thickness of the free layer. For our device simulations we consider $\xi = 130$ (fJVm$^{-1}$) and $t_F = 1\ nm$, which corresponds to voltage = 0.45 V. In Fig. 3(a) we show how the output voltage of the MTJ evolves in time in the presence of voltage pulses/spikes, with increasing input spike frequency.

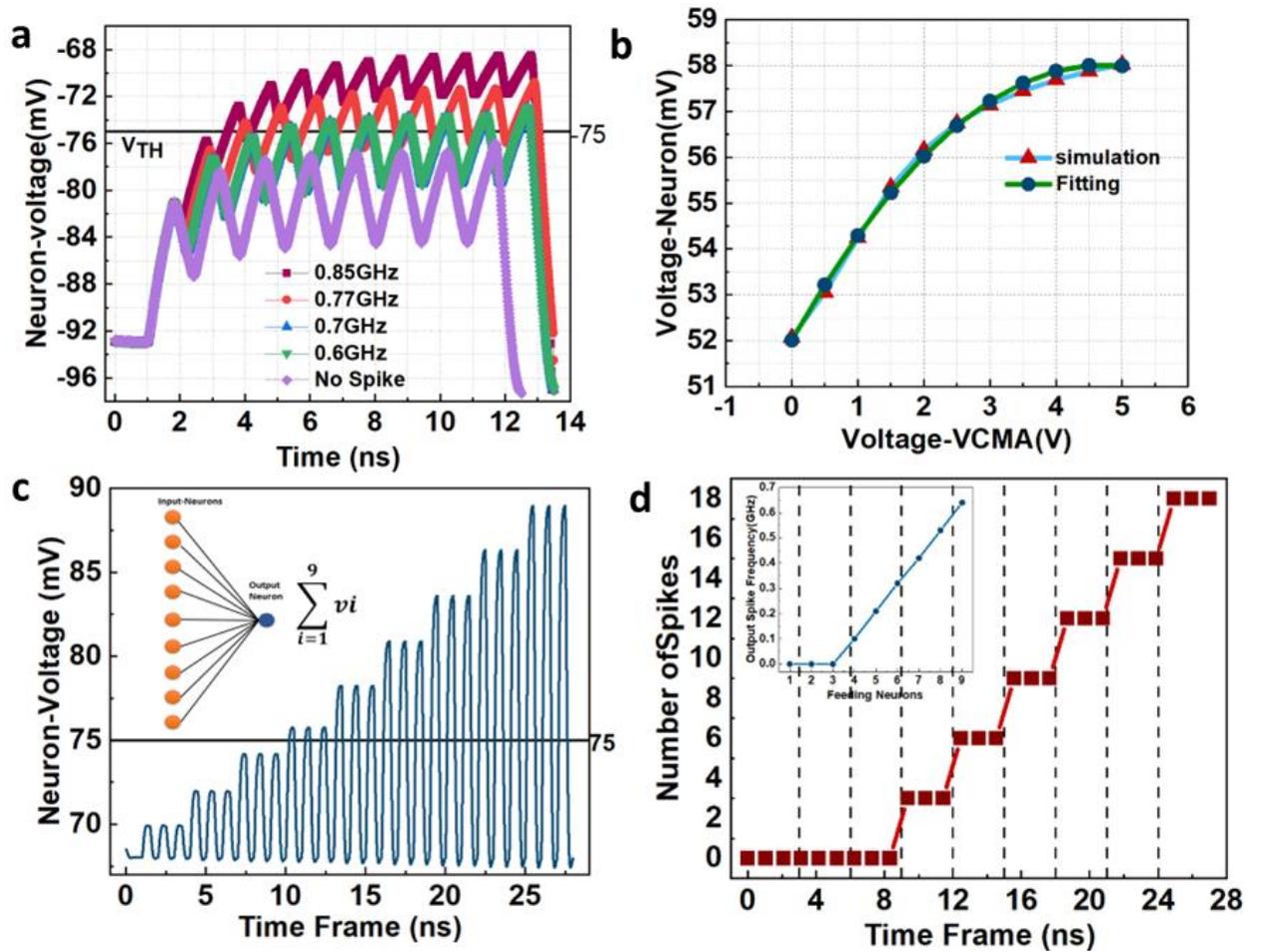

Fig. 3(a) Skyrmion output voltage mimics the leaky integrate and fire neuron behavior for different input spike frequency. (b) Input voltage vs output voltage simulation and analytical fitting (c) 9 input neurons feeding one output neurons in different time frames. (d) Output spike rate in different time frame Inset: Output neuron spikes frequency increases as more input neurons start feeding it.

The MTJ magnetization and voltage switching characteristics mimics the leaky integrate and fire (LIF) neuron. With increase in pre-neuron spike frequency the value of normalized magnetizations of post-neuron increases from -97 mV to -68mV.

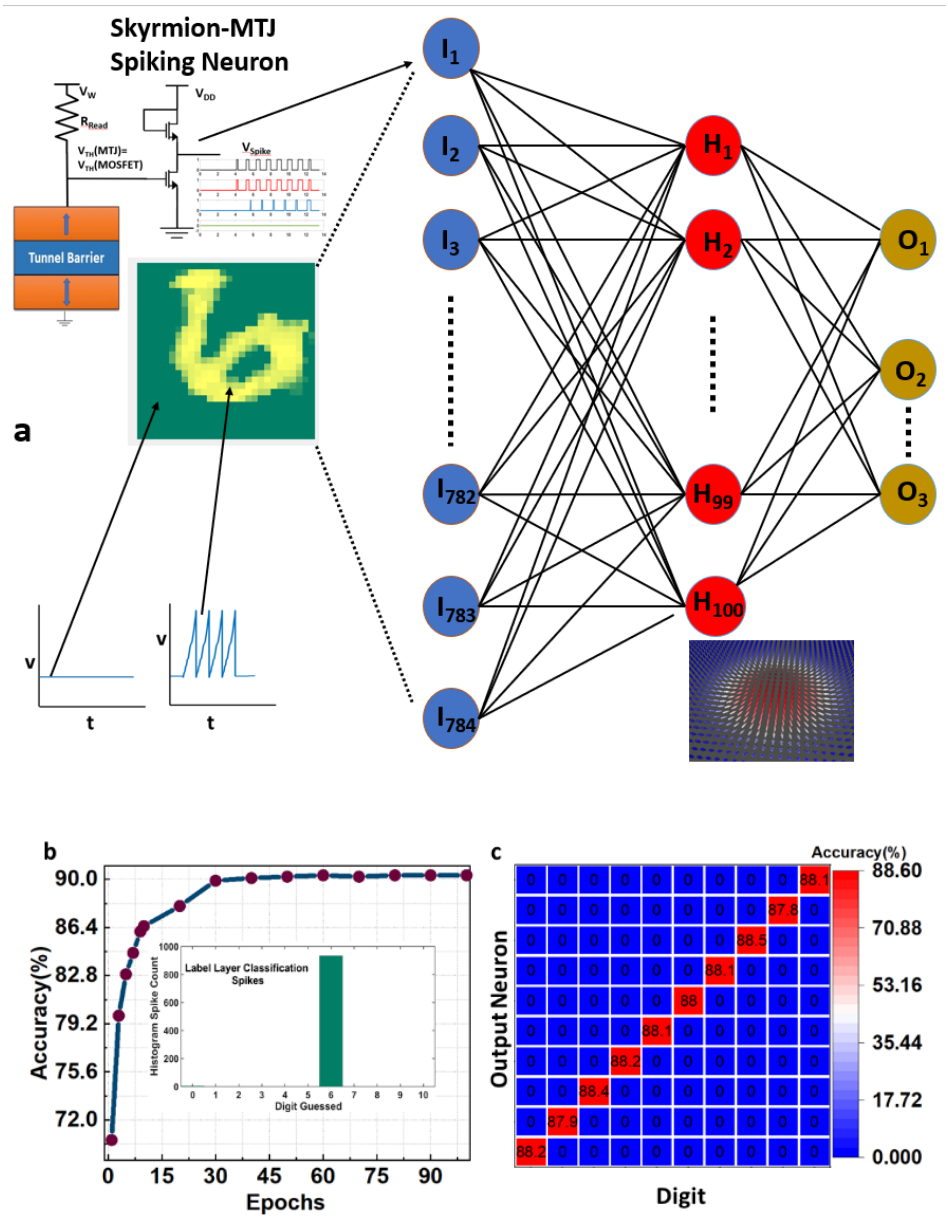

Fig. 4(a)The training and testing of MNIST handwritten digit-(6). The DBN is trained on RBM architecture (b) SNN accuracy increases with the number of epochs and we obtain best accuracy of 90%. Inset: spike count of output neuron -6. (c) Confusion matrix for MNIST digit recognition, x-axis represents the test digit y-axis shows the accuracy of output neuron.

We keep the normalized threshold voltage $V_{TH}$ at -75mV and it can be observed that both output spike rate and width of spikes increases with increasing input(pre-neuron) spike rate. Further, the neuron frequency response increases as more input neurons start feeding it. In Fig. 3(b) we show the input output voltage characteristics of the MTJ neuron. The neuron output behavior is quadratic with respect to input voltage and the output is analytically fitted by following expression.

$$V_{NEURON} = \beta V_k^2 + \gamma V_k + 64 \qquad (6)$$

where, $V_k$ is the VCMA voltage, $\beta = -0.2757$ (mV)$^{-1}$

and $\gamma = 2.55$ (unit less)Along with the input spike frequency, the output spike frequency can be also tuned by gate voltage. As shown in Fig. 3(c) nine input neurons feed a single output neuron. The input voltage peak increases in increasing time frames. As shown in Fig. 3(d), in different time frames, with more input neurons getting activated the VCMA gate voltage increases and correspondingly output spike frequency increases as shown in inset Fig. 3(b). Furthermore, using the SNN toolbox (edbn-master) in MATLAB toolbox. we evaluate the neuromorphic computing capabilities, based on the proposed device we implement a SNN with 784 input neurons, 100 hidden neurons and 10 output neurons (MATLAB) (see Fig. 4a). The edbn toolbox maps deep belief network (DBN) weights in restricted Boltzmann machine (RBM) architecture to the SNN. We show the pattern classification on the MNSIT dataset using the parameters corresponding to skyrmion LIF neuron. The parameters such as $V_{rest}$, $V_{reset}$, time constant $\tau_m$ and threshold voltage $V_{th}$ for the neuron are taken from Fig. 3(a) $V_{rest}$ (-90mV), $V_{reset}$ (-50mV) and $V_{th}$ (-75mV) *and* $\tau_m$ = *1ns*. We obtain a maximum accuracy of around 90% for 40 epochs as shown in Fig. 4(b). The inset in Fig. 4(b) shows the spike count of output neuron 6 is maximum among all the digits. Fig. 4(c) represents the confusion matrix corresponding to all 10 MNIST digits, we observe the network was able to classify all digits with accuracy ~88%. Note: In-situ training was performed using supervised learning approach.

CONCLUSION

We propose a high-integration density and energy-efficient magnetic skyrmion-device structure, based on a CoFeB/MgO/ CoFeB/Ta MTJ, that can behave like a LIF neuron when pulsed voltage stresses are applied. The confinement and voltage control offers a possible path toward utilizing skyrmions in logic and memory units. In addition, the skyrmion hall effect (which is a major bottleneck for skyrmion racetrack memory) is avoided. The interaction between neighboring skyrmions is also minimized by geometry and stress engineering. The proposed neuron model is tested on MNIST data and shows 90% pattern classification efficiency.